# Enhanced Bayesian Model Updating with Incomplete Modal Information Using Parallel, Interactive and Adaptive Markov Chains

K. Zhou

Postdoctoral Researcher

J. Tang[†]

Professor

Department of Mechanical Engineering

University of Connecticut

191 Auditorium Road, Unit 3139

Storrs, CT 06269

USA

Phone: (860) 486-5911, Email: jiong.tang@uconn.edu

**Submitted to:**

---

[†] Corresponding author

# Enhanced Bayesian Model Updating with Incomplete Modal Information Using Parallel, Interactive and Adaptive Markov Chains

K. Zhou and J. Tang†

Department of Mechanical Engineering

University of Connecticut

Storrs, CT 06269, USA

Phone: +1 (860) 486-5911, Email: jiong.tang@uconn.edu

## ABSTRACT

Finite element model updating is challenging because 1) the problem is oftentimes underdetermined while the measurements are limited and/or incomplete; 2) many combinations of parameters may yield responses that are similar with respect to actual measurements; and 3) uncertainties inevitably exist. The aim of this research is to leverage upon computational intelligence through statistical inference to facilitate an enhanced, probabilistic finite element model updating using incomplete modal response measurement. This new framework is built upon efficient inverse identification through optimization, whereas Bayesian inference is employed to account for the effect of uncertainties. To overcome the computational cost barrier, we adopt Markov chain Monte Carlo (MCMC) to characterize the target function/distribution. Instead of using single Markov chain in conventional Bayesian approach, we develop a new sampling theory with multiple parallel, interactive and adaptive Markov chains and incorporate into Bayesian inference. This can harness the collective power of these Markov chains to realize the concurrent search of multiple local optima. The number of required Markov chains and their respective initial model parameters are automatically determined via Monte Carlo simulation-based sample pre-screening followed by *K*-means clustering analysis. These enhancements can effectively address the aforementioned challenges in finite element model updating. The validity of this framework is systematically demonstrated through case studies.

## 1. Introduction

Finite element (FE) method nowadays is pervasive in the analysis of mechanical structures [1-4]. A critical step in FE analysis is model updating in which certain modeling parameters are identified/updated to minimize the discrepancy between FE-based response prediction and the actual measurement. Indeed, in recent years, there have been growing interests in FE model updating using vibration response data to facilitate a series of inter-related applications, such as damage identification, design optimization, and structural control, etc [2-4]. In dynamic systems, FE model updating generally is conducted by employing the responses in either time or frequency domain [5-7]. As one type of inherent characteristics of a structure, mode shapes and their curvatures have been employed owing to their capability of reflecting local structural property variation [8-10]. Most previous research efforts in this regard, however, have been conducted toward the deterministic case, i.e., the FE model is deterministic whereas all the information involved including measurements is also deterministic. In reality, the baseline finite element model to be updated is subject to numerical modeling error, and many parameters involved in the model are intrinsically uncertain due to manufacturing tolerance and measurement noise/error. A deterministic model updating procedure cannot effectively address such uncertainties.

---

† Corresponding author

As can be seen, model updating should be conducted in the probabilistic sense, i.e., treating model parameters to be updated as random variables with mean and variance. This can reveal the underlying properties of structures under uncertainties and variations. There have been some probabilistic approaches developed to investigate the parametric estimation in the presence of uncertainties. For example, Moaveni et al [11] implemented sensitivity-based finite element model updating for damage identification, in which the uncertainty level that affects the identification result is quantified through analysis-of-variance (ANOVA) and meta-modeling. Khodaparast et al [12] used kriging predictor to conduct interval model updating where irreducible uncertainty was considered. Bayesian inference naturally appears to be one of the most popular approaches, in which a probabilistic model is established to correct the prior beliefs based on the evidences [13]. It starts from characterizing the concerned model parameters in the form of probability density function (PDF) based upon the prior knowledge. This specific PDF is referred to as prior PDF or hypothesis in Bayes' rule. The actual response measurement is treated as evidence, and can be incorporated to update the prior PDF into the so called posterior PDF, based upon which the best model parameters can be identified. Bayesian inference not only can avoid the direct inversion for parameter estimation required in some sensitivity-based methods that may introduce some numerical issues, but also can directly incorporate various sources of uncertainties into model updating procedure [14]. Owing to its intrinsic advantages, there have been considerable successes in utilizing Bayesian inference to solve a variety of engineering problems [4, 15-18]. Additionally, as the training scheme of meta-models, which are essentially implicit statistical regression models, is generally established upon Bayesian inference, exploring the potential close-form of optimization objective function under Bayesian framework may benefit meta-model training efficiency [19].

It is worth noting that, while Bayesian inference can enable probabilistic FE model identification under uncertainties, currently its application to complex structures is subject to certain limitations. One significant limitation lies in the huge computational cost of brute force Monte Carlo simulations of repeated finite element analyses. The computational cost will become intractable especially when the number of model parameters to be updated increases. Indeed, as the number of parameters to be updated increases, the search space dimension increases which requires a very large number of FE simulation runs in order to identify the updating result. The usual treatment to alleviate the computational cost is either to develop first principle-based order-reduction model [20] or data-based surrogate model [21] to replace the original, large-scale finite element model. The first category of methods is inevitably subject to model truncation error due to order-reduction. Such error may become considerable when compared with the discrepancy between the actual measurement and finite element model prediction. In the second category of methods, it is difficult to rigorously determine the size of dataset needed. In certain cases, it is even difficult to decide how to select dataset from FE simulations to train/establish a surrogate model to accurately approximate the original FE model. On the other hand, improved sampling techniques, which may reduce the computational cost through reducing the number of FE simulation runs, have been extensively investigated. One popular approach is the Markov chain Monte Carlo (MCMC) method, which can be seamlessly integrated into Bayesian inference-based optimization [22-25]. It comprises a class of algorithms e.g., Metropolis-Hastings (MH) [26], Gibbs sampling (GS) [27] and importance sampling [28] to enable efficient sampling from an unknown target distribution/function. The constructed Markov chain hence is deemed as an equilibrium distribution of target distribution/function [29]. The number of samples in Markov chain is considerably smaller. Among these algorithms, MH-based MCMC is the most common method used in the application of FE model updating [26, 30]. When MH-based MCMC method is applied, a proposal distribution is formulated to guide the sample generation over the entire parametric space. Generally, the variance of the proposal distribution is set as constant in the course of chain evolution.

The underlying idea of MCMC can lead to an accelerated approximation of target distribution. Here the target distribution essentially represents the actual objective surface in model updating. In this research, the actual objective surface is defined as the error surface between the FE prediction and the measurement with respect to the model parameter samples. It is worth noting that, in almost all practical situations, the number of sensors is limited and much smaller than the number of degrees of freedom (DOFs) in the structural model, and generally only the dynamic responses within the lower-order frequency ranges can be realistically measured. Therefore, the measurement acquired is incomplete. In many cases the inverse analysis-based model updating problem is underdetermined. Consequently, the objective surface may become very complex, exhibiting many local optima. Conventional Bayesian model updating is performed with single MCMC, which is only capable of converging to one optimum. This optimum may very well be a local one. The proposal distribution in MCMC usually has fixed variance which may further increase the chance of being trapped in local optimum. Several approaches have been attempted to address this issue. For example, Liang et al [31] proposed a Markov chain Monte Carlo method with adaptive proposal distribution for performance enhancement. Ji and Schmidler [32] formulated a mixture proposal distribution which can adapt to samples from multimodal target distribution, and demonstrated improved approximation. Recently, Lam et al [33] developed a multiple parallel MCMC-based Bayesian model updating approach to ensure the accuracy of updating results.

The objective of this research is to fundamentally address the challenges in Bayesian model updating in FE analysis, i.e., underdetermined problem with complex objective surface. Specifically, we enhance the Bayesian model updating framework with the integration of multiple parallel, interactive and adaptive Markov chains. We not only maintain the parallel scheme of Markov chains [33], but also enable all Markov chains to evolve in an interactive manner. The redundant Markov chains that yield the same local optima with others will be suspended in order to alleviate the computational cost. Meanwhile, an automatic analysis procedure is employed to adaptively determine the number of Markov chains and related initial parameters to be executed. As will be demonstrated in this research, these strategies can take full advantage of the characteristics of dynamic responses utilized in FE model updating such as incomplete mode shape measurements to unleash the potential of physics informed statistical inference. The rest of this paper is organized as follows. In Section 2, the general formulation of finite element model updating using incomplete mode shape information is outlined first, followed by an overview of traditional Bayesian inference-based model updating framework integrated with single Markov Chain which serves as the baseline. Subsequently, the enhanced framework built upon multiple parallel, interactive Markov chains is then presented. Section 3 provides implementation details and systematic case studies on a benchmark structure to demonstrate the proposed methodology and illustrate the performance improvement. Section 4 gives the concluding remarks.

## 2. FE Model Updating with Enhanced Bayesian Framework: Algorithm Development

This section presents the formulation of FE model updating with enhanced Bayesian framework. We start from the problem formulation of FE model updating using (incomplete) mode shape measurements. It is followed by the outline of conventional Bayesian updating. The new framework is then established where the enhancement through incorporating multiple Markov chains is highlighted.

### *2.1. Problem formulation of FE model updating using incomplete modal response measurement*

Finite element (FE) model updating is a widely used procedure to identify model parameters based on measurement. The response prediction from a baseline finite element model, under given (sampled) parameters is compared with the measurement from actual structure to facilitate the entire updating process. The FE-based dynamic equation of motion can be expressed as

$$\mathbf{M\ddot{x}} + \mathbf{C\dot{x}} + \mathbf{Kx} = \mathbf{F} \qquad (1)$$

where $\mathbf{M}$, $\mathbf{C}$, and $\mathbf{K}$ are the mass, damping, and stiffness matrices, respectively. We assume the structure has $N$ DOFs. All the system matrices are of dimension $N \times N$. $\mathbf{x}$ is the $N$-dimensional displacement vector. Let the structure be subject to light and proportional damping. The damping matrix thus is dependent on the mass and stiffness matrices. Without loss of generality, in what follows we are only concerned about the updating of stiffness matrix. In general the updating of both the mass and stiffness matrices can be formulated similarly.

In practice, owing to the high dimensionality of FE model, it is impossible to update all elemental stiffness and mass matrices. Commonly, we assume that variations of model parameters (i.e., parameters to be updated) only occur at $n$ ($n \ll N$) elements or segments in the FE model. Let $\mathbf{K}_i$ denote the nominal stiffness matrix of the $i$-th segment in the original, baseline FE model. The global stiffness matrix that is subject to variation thus can be expressed as [34]

$$\hat{\mathbf{K}} = \sum_{i=1}^{n} \mathbf{K}_i (1 - \alpha_i) \qquad (2)$$

where $\alpha_i$ indicates the stiffness variation coefficient of the $i$-th segment. In this research, we let $\alpha_i$ fall into $[0, 1]$. The coefficient vector, $\boldsymbol{\alpha} = [\alpha_1, \alpha_2 ... \alpha_i ... \alpha_n]$, represents the unknown and random stiffness variations to be updated. It is worth noting that the aforementioned formulation can be directly applied to damage identification case. In such case, the original, baseline FE model corresponds to the healthy structure, and $\boldsymbol{\alpha} = [\alpha_1, \alpha_2 ... \alpha_i ... \alpha_n]$ then represents the unknown stiffness reductions caused by damage. In other words, Equation (2) can be used in both model updating and damage identification.

The stochastic FE model of the actual structure, that is perturbed by the unknown parameters to be updated, thus becomes

$$\hat{\mathbf{M}}\ddot{\mathbf{x}} + \hat{\mathbf{C}}(\boldsymbol{\alpha})\dot{\mathbf{x}} + \hat{\mathbf{K}}(\boldsymbol{\alpha})\mathbf{x} = \mathbf{F} \qquad (3)$$

Here $\hat{\mathbf{M}} = \mathbf{M}$ which remains invariant, and the stiffness and damping matrices are affected by $\boldsymbol{\alpha}$. The modal information of this system is governed by the following eigenvalue problem,

$$[\hat{\mathbf{K}}(\boldsymbol{\alpha}) - (2\pi\hat{\omega}_i)^2 \hat{\mathbf{M}}(\boldsymbol{\alpha})]\hat{\boldsymbol{\psi}}_i = \mathbf{0} \qquad (4)$$

The $i$-th natural frequency and the corresponding mode shape are denoted as, $\hat{\omega}_i$ and $\hat{\boldsymbol{\psi}}_i$, respectively. They are both functions of $\boldsymbol{\alpha}$.

In model updating practice, usually the first $q$ ($q \ll N$) lower-order natural frequencies and mode shapes can be experimentally extracted. In actual data acquisition, only a small number of $s$ ($s \ll N$) sensors can be employed at the corresponding DOFs for measurement [10]. The modal information acquired thus is incomplete and limited. For notation convenience, we let ($\bar{\boldsymbol{\omega}}_{1\times q}$, $\bar{\boldsymbol{\psi}}_{s\times q}$) denote the measurement information consisting of first $q$ natural frequencies and an $s\times q$ matrix of the collection of corresponding mode shape vectors. Hereafter the subscripts of these variables indicate their dimensions. In each mode shape vector, only the modal amplitudes at $s$ DOFs are measured. Similarly, we let ($\hat{\boldsymbol{\omega}}_{1\times q}$, $\hat{\boldsymbol{\psi}}_{s\times q}$) denote the information simulated/predicted from the FE model (Equation (4)). Our goal is to identify the set of uncertain parameters, $\boldsymbol{\alpha}$, such that the difference between ($\bar{\boldsymbol{\omega}}_{1\times q}$, $\bar{\boldsymbol{\psi}}_{s\times q}$) and ($\hat{\boldsymbol{\omega}}_{1\times q}$, $\hat{\boldsymbol{\psi}}_{s\times q}$) is minimized. While the mathematical details will be provided subsequently, here we generically express the differences of natural frequencies and mode shapes as $\kappa(\bar{\boldsymbol{\omega}}_{1\times q}, \hat{\boldsymbol{\omega}}_{1\times q}(\boldsymbol{\alpha}))$ and $\upsilon(\bar{\boldsymbol{\psi}}_{s\times q}, \hat{\boldsymbol{\psi}}_{s\times q}(\boldsymbol{\alpha}))$, respectively, which are to be minimized. As mentioned, in this research the methodology to be developed will incorporate various uncertainties and measurement noise. Therefore, $\kappa(\bar{\boldsymbol{\omega}}_{1\times q}, \hat{\boldsymbol{\omega}}_{1\times q}(\boldsymbol{\alpha}))$ and $\upsilon(\bar{\boldsymbol{\psi}}_{s\times q}, \hat{\boldsymbol{\psi}}_{s\times q}(\boldsymbol{\alpha}))$ are no longer deterministic. They are instead probabilistic. The model parameters to be updated, $\boldsymbol{\alpha}$, will be identified in a probabilistic manner accordingly.

## 2.2. Conventional Bayesian model updating with MCMC
### 2.2.1. Bayesian inference

The objective of this research is to develop a new framework to update, probabilistically, the FE model based on measurement information. The underlying idea of Bayesian inference is to update the probability of hypothesis as more evidences become available, which fits the research objective. It is well known that Bayesian inference is established upon the Bayes' rule which is fully represented as [13]

$$p(\boldsymbol{\theta}|\boldsymbol{\Omega}) = \frac{p(\boldsymbol{\Omega}|\boldsymbol{\theta})p(\boldsymbol{\theta})}{\int p(\boldsymbol{\Omega}|\boldsymbol{\theta})p(\boldsymbol{\theta})d\boldsymbol{\theta}} \tag{5}$$

In the context of FE model updating, the hypothesis $\boldsymbol{\theta}$ is interpreted as the vector of model parameters (i.e., $\boldsymbol{\alpha}$ in Equation (2)) to be updated. The evidence, $\boldsymbol{\Omega}$, is the modal information difference (between model prediction and actual measurement), i.e., $\kappa(\bar{\boldsymbol{\omega}}_{1\times q}, \hat{\boldsymbol{\omega}}_{1\times q}(\boldsymbol{\alpha}))$ and $\upsilon(\bar{\boldsymbol{\psi}}_{s\times q}, \hat{\boldsymbol{\psi}}_{s\times q}(\boldsymbol{\alpha}))$. Hereafter to simplify the notations we refer to the differences of natural frequencies and mode shapes as $\kappa(\boldsymbol{\alpha})$ and $\upsilon(\boldsymbol{\alpha})$. We thus re-write the above equation as

$$p(\boldsymbol{\alpha}|\kappa(\boldsymbol{\alpha}), \upsilon(\boldsymbol{\alpha})) = \frac{p(\kappa(\boldsymbol{\alpha}), \upsilon(\boldsymbol{\alpha})|\boldsymbol{\alpha})p(\boldsymbol{\alpha})}{\int p(\kappa(\boldsymbol{\alpha}), \upsilon(\boldsymbol{\alpha})|\boldsymbol{\alpha})p(\boldsymbol{\alpha})d\boldsymbol{\alpha}} \tag{6}$$

The prior PDF $p(\boldsymbol{\alpha})$ is an arbitrary distribution of $\boldsymbol{\alpha}$ initially created based on prior knowledge. Without explicit understanding of the target problem, this term can be simply defined as a standard statistical distribution, such as normal or uniform distribution. The likelihood PDF $p(\kappa(\boldsymbol{\alpha}), \upsilon(\boldsymbol{\alpha})|\boldsymbol{\alpha})$ aims at probabilistically assessing the agreement between the measurement and the corresponding modal information prediction from the model. The posterior PDF $p(\boldsymbol{\alpha}|\kappa(\boldsymbol{\alpha}), \upsilon(\boldsymbol{\alpha}))$ is the resultant distribution of $\boldsymbol{\alpha}$ conditioned on the prior PDF and the measurement. Since the marginal likelihood in the denominator essentially is a normalization constant, the posterior PDF is proportional to the numerator, i.e., $p(\boldsymbol{\alpha}|\kappa(\boldsymbol{\alpha}), \upsilon(\boldsymbol{\alpha})) \propto p(\kappa(\boldsymbol{\alpha}), \upsilon(\boldsymbol{\alpha})|\boldsymbol{\alpha}) p(\boldsymbol{\alpha})$. It is worth noting that the posterior PDF here is considered as the optimization objective to guide the model updating/optimization process, in which the best parametric combination, $\tilde{\boldsymbol{\alpha}}$, can be identified with the highest $p(\tilde{\boldsymbol{\alpha}}|\kappa(\boldsymbol{\alpha}), \upsilon(\boldsymbol{\alpha}))$.

To implement the Bayesian inference, we will need to build the probabilistic relationship between the measurement and corresponding modal information prediction in the presence of uncertainties. In this research, measurement and modeling errors are considered as Gaussian noise. The likelihood PDF that takes into account the effect of these errors thus is subject to a multivariate normal distribution. The specific formulation of likelihood PDF is dependent on the explicit forms of $\kappa(\boldsymbol{\alpha})$ and $\upsilon(\boldsymbol{\alpha})$, which will be discussed later.

*2.2.2. Integration of MCMC for expedited optimization*

In order to identify the model parameters to be updated, Bayesian inference-based optimization will be conducted which requires Monte Carlo simulation with repeated FE analyses. For a practical structure, each FE analysis run will take certain computational cost. A brute force Monte Carlo simulation would be computationally prohibitive, as a very large number of parametric combinations need to be substituted into this procedure to construct a credible posterior PDF. The computational issue will be further compounded when high-dimensional parametric set, i.e., large *n* in Equation (2), is involved. A common solution is to adopt Markov chain Monte Carlo (MCMC) to replace the conventional Monte Carlo to facilitate efficient model updating analysis under uncertainties.

The fundamental idea of MCMC is that it can generate a stationary chain with a significantly reduced number of model parameter samples that are used to approximate the target distribution [23, 29, 35]. The approximated distribution is capable of interpreting the best parametric combination to be identified in a probabilistic sense. There are several algorithms that can be utilized to execute MCMC. Here we adopt the Metropolis-Hastings (MH) MCMC with pseudo code shown below.

Pesudo code of Metropolis-Hastings MCMC

With $\boldsymbol{\alpha}_t^*$ at time *t*, the aim is to generate the next chain value $\boldsymbol{\alpha}_{t+1}^*$.

1. Proposal step: Sample "Candidate" *z* from the proposal distribution $\mathbf{Z} \sim q(\mathbf{z}|\boldsymbol{\alpha}_t^*)$. Proposal distribution

usually is selected as a symmetrical distribution, e.g., normal distribution.

2. Acceptance step: With probability $\beta(\mathbf{Z},\mathbf{\alpha}_t^*) = \min(1, \frac{p(\mathbf{\alpha}_t^*|\kappa(\mathbf{\alpha}_t^*),\upsilon(\mathbf{\alpha}_t^*))q(\mathbf{Z}|\mathbf{\alpha}_t^*)}{p(\mathbf{Z}|\kappa(\mathbf{Z}),\upsilon(\mathbf{Z}))q(\mathbf{\alpha}_t^*|\mathbf{Z})})$. Due to the symmetry of proposal distribution, $q(\mathbf{Z}|\mathbf{\alpha}_t^*)=q(\mathbf{\alpha}_t^*|\mathbf{Z})$.

Generate $\mu$ from a uniform (0, 1) distribution:

if $\beta(\mathbf{Z},\mathbf{\alpha}_t^*) > \mu$, we set $\mathbf{\alpha}_{t+1}^* = \mathbf{Z}$ (i.e., acceptance)

else $\mathbf{\alpha}_{t+1}^* = \mathbf{\alpha}_t^*$ (i.e., rejection).

Note: $q(\mathbf{z}|\mathbf{\alpha}_t^*)$ is the proposal distribution of $z$ depending on the deterministic parameter $\mathbf{\alpha}_t^*$.

In MH MCMC, 'Metropolis criterion' is strictly followed to determine whether the newly generated sample is retained or discarded. Through executing the analysis of MCMC, we can obtain the Markov chain containing all accepted model parameter samples and their posterior PDF values. The reduced number of samples in MCMC leads to significant reduction of computational cost. The accepted model parameter samples in Markov chain can be further used to estimate/approximate the posterior PDF for probabilistic parameter updating. The conventional single MCMC-based Bayesian model updating framework introduced in this Section is shown in Figure 1.

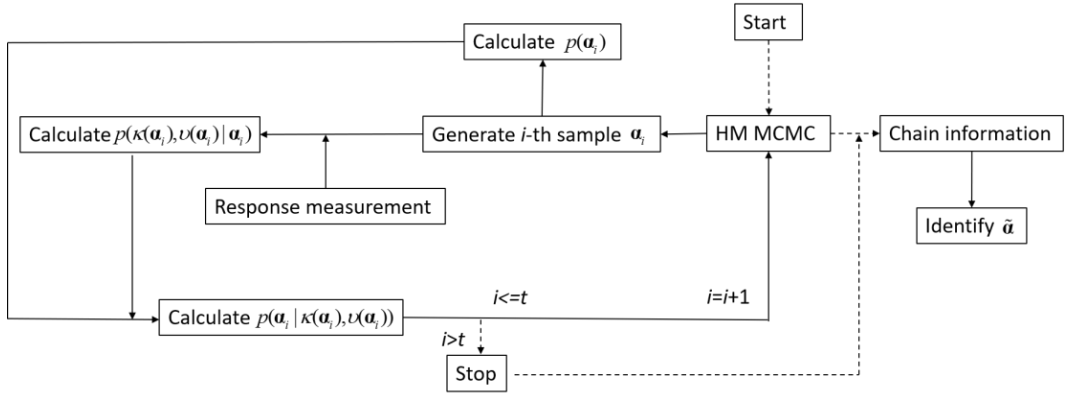

Figure 1. Conventional Bayesian model updating.

## 2.3. Enhanced Bayesian model updating framework with integration of multiple parallel, interactive and adaptive Markov chains

In model updating using vibration response, generally only the lower order modal information can be realistically measured. Moreover, as the number of sensors is limited, the mode shape information is incomplete. When the number of model parameters to be updated is large, the model updating problem is oftentimes underdetermined. Moreover, the objective surface in inverse identification oriented
8

optimization is very complex, exhibiting many local optima. Conventional Bayesian model updating performed with single MCMC which is outlined in the Section 2.2 is only capable of converging to one optimum which may not likely be the best result. To tackle this challenge, in what follows we develop the key component in the enhanced model updating framework, i.e., the parallel, interactive and adaptive Markov chains, which aims at adequately approximating the target distribution over the entire space of model parameters to be updated.

We first want to ensure capturing as many local optima as possible, and thus adopt a parallel scheme of Markov chains, similar to what's suggested by literature [33]. Generally, to increase updating accuracy, the number of Markov chains, i.e., $m$, is suggested to be large especially for cases with complicated objective surfaces to be characterized. The $m$ resulting Markov chains can be utilized to identify the probabilistic solution. Building upon this, we include an adaptive scheme for varying proposal distribution width in order to improve the performance of MCMC [31]. Specifically, we assign a small width $\gamma$ for proposal distribution in the beginning, since different Markov chains are designated to search parameters at respective local areas. A small width thus benefits the dense search process. During the MCMC evolution, distribution width varies following the specified rule. For instance, if the number of consecutive sample rejections exceeds a specified threshold, we increase the distribution width to enlarge the search space which avoids trapping by local optima. Once the sample is accepted, the default small width $\gamma$ is then restored.

There are, however, remaining issues. The first issue is that the selection of $m$ Markov chains. In earlier investigations, $m$ is manually selected according to the configuration of computational platform, such as the number of processors. In particular, the initial parameters of $m$ chains are either randomly generated based upon a pre-specified statistical distribution or grid-discretized to uniformly cover the entire parametric space. These procedures may not yield optimal initial parameters, which thus slows down the convergence of Markov chains. The second issue is that, while these $m$ Markov chains evolve independently in parallel, it is not guaranteed that all of them will finally converge to $m$ different, well-separated local optima. Additionally, if some Markov chains already converge to the same local optima and still are allowed for continuous evolution, computational resource will be occupied unnecessarily. In this research we aim at developing a new Bayesian model updating framework using incomplete modal information. While the mathematical details will be presented in the next section, the treatments to address the aforementioned issues are induced as follows.

We establish a sequential procedure for the $m$ parallel Markov chains (Figure 2). We construct a uniform distribution covering the entire parametric space, based on which we randomly generate $w$ model parameter samples. Without prior knowledge, we perform the Monte Carlo simulation to calculate the objective values (i.e., posterior PDF values) of these parameter samples following Equation (6). We then sort the objective values and find the *r* best model parameter samples with higher objective values. Following this, we utilize the *K*-means clustering analysis to take advantage of the sorted samples. The *K*-means clustering analysis can partition data into different clusters with the nearest mean [36]. The clustering process is implemented by iteratively updating the means of obtained clusters until the convergence criterion is met. The distance between two model parameter samples are represented by the Euclidean distance of the associated spatial coordinates [37]. We will implement *K*-means clustering analysis based on *w* model parameter samples to determine the maximum number of clusters *m* ( $m \leq r$ ) that can satisfy the distance constraint, i.e., the distance between any two clusters should exceed the specified minimal distance. We further use *m* centers of *m* clusters as initial model parameters to execute the *m* parallel Markov chain evolutions.

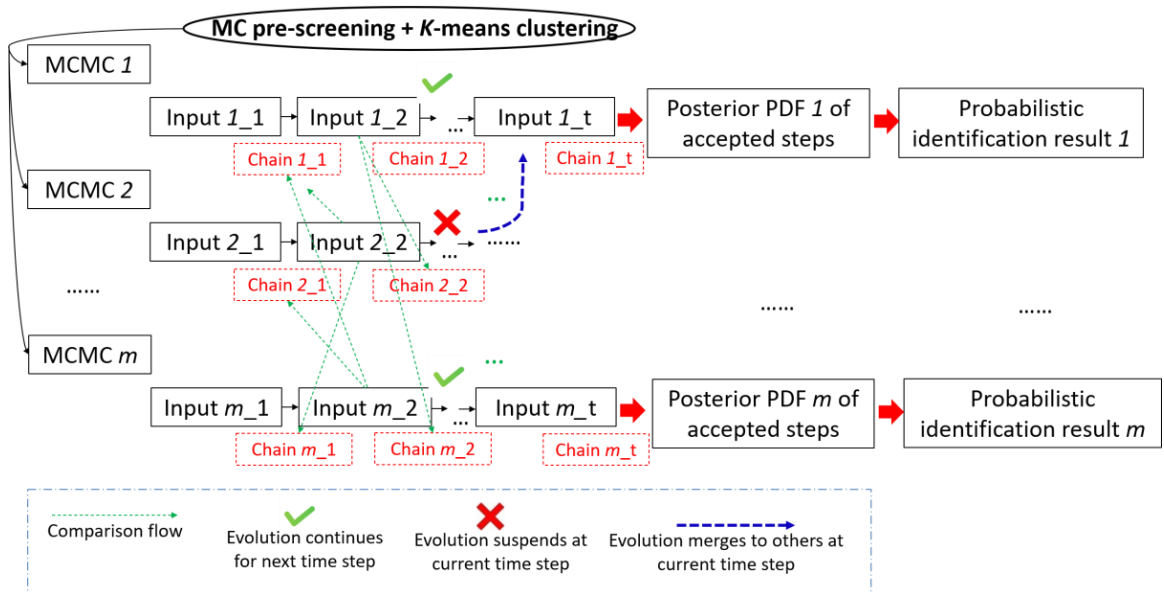

Figure 2. Schematic illustration of integration of *parallel*, *interactive* and *adaptive* Markov chains.

To address the second issue in the parallel Markov chains, we integrate a procedure for the merge check of Markov chains during evolutions. At each time step, currently accepted model parameter sample of one certain MCMC will be compared with all accepted model parameter samples archived in other Markov chains. When it occurs that the Euclidean distance of two compared model parameter samples is smaller than a threshold, one MCMC is suspended and will not continue to evolve in the next time step. This allows the interaction of all Markov chains. The final number of the resulting Markov chains is *z*, where $z \leq m$. Correspondingly, *z* solutions, i.e., *z* unique local optima, will be identified through the chain information obtained. There indeed have been multiple criteria suggested for terminating the MCMC evolution. One popular criterion is dependent on the convergence examination, in which the auto-correlation degree of Markov chain is iteratively assessed as the evolution proceeds [38]. A threshold of auto-correlation degree will be prescribed to decide whether the evolution can be terminated. Another criterion is to directly assign a maximum iteration number for MCMC evolution [10]. In this research, for the sake of illustration we adopt the latter one. When a pre-specified iteration number, i.e., $t$ of Markov chain, is reached, the whole process will be terminated. The earlier termination also allows to ensure the computational efficiency when the specified number of consecutive rejected steps, i.e., $u$ is reached. The eventually survived Markov chains can offer the information to extract the probabilistic updating results.

This enhanced framework is schematically illustrated in Figure 2. To maintain the stationary property of MCMC, first a few accepted samples in the transition phase, the so-called burn-in period, need to be scrapped [39]. Hence, the burn-in length ratio $\kappa$ should be defined in the analysis.

## 3. New Framework Implementation Details and Case Demonstrations

This section presents the mathematical details of the enhanced Bayesian model updating framework through implementing case investigations. In order to validate the effectiveness and generality of the new approach, we practice the algorithmic implementation to two different, representative scenarios, i.e., identifying boundary conditions of a dome structure and updating material properties of a plate structure with high FE mesh density. In both cases, limited and incomplete modal information will be used to facilitate model updating. In this research we use simulated data in lieu of experimental data to facilitate model updating for algorithm investigation. This allows interested readers to reproduce the results for examination. Moreover, since the 'ground truth' is known in the simulated cases, the effectiveness and accuracy of the new framework can be thoroughly investigated, especially in the presence of many local optima in Bayesian inference based optimization.

### 3.1. Implementation scenario 1: Boundary condition updating of a dome structure
#### 3.1.1. Model updating problem setup

We choose a dome-type structure as the first implementation scenario, as it is representative of many civil infrastructures. The configuration and geometric parameters are shown in Figure 3. It is made of with homogeneous material with Young's modulus $2.06 \times 10^{11}$ Pa, mass density $7.85 \times 10^3$ kg/m$^3$, and Poisson's ratio $0.3$. The finite element model of this structure is built with the beam element, containing a total 108 of nodes (each node with 3 translational DOFs). The cross section of beam element is rectangular with area $0.0168$ m$^2$ (cross section length and width: 0.21 m and 0.08 m, respectively). In this dome structure, 18 nodes at the bottom layer are imposed with boundary conditions, i.e., being fixed to ground. We assume in this case 14 out of 18 nodes are indeed completely fixed (i.e., no displacement at all 3 directions). And we assume the other 4 nodes are not fixed ideally, with stiffness parameters to be updated based on simulated measurement information. To a large extent, updating boundary conditions essentially is to calibrate the stiffness at boundary DOFs. To facilitate such analysis, we employ 3 spring elements that are aligned with the 3 principal directions of each node and connected with the ground to emulate the boundary conditions (Figure 4). If a boundary DOF is completely fixed, the corresponding spring stiffness is theoretically infinity. On the other hand, if a boundary DOF is not completely fixed, a finite spring stiffness value will need to be identified.

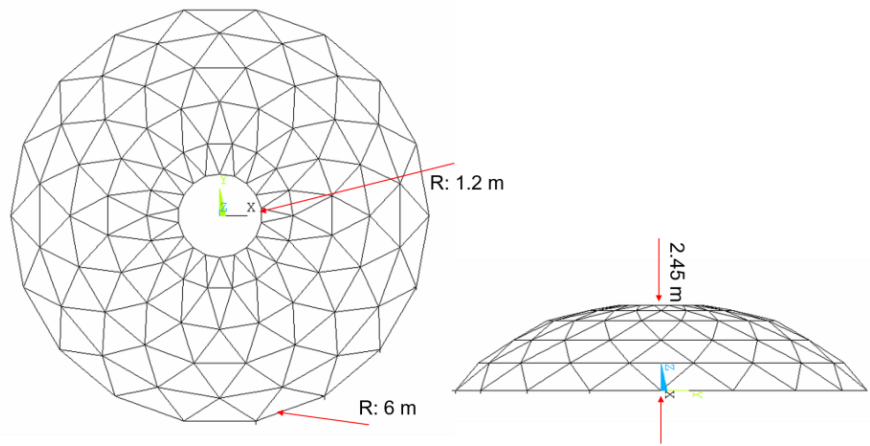

Figure 3. The geometry of the dome structure. Outer radius: 6m; inner radius: 1.2m; height: 2.45m.

Without loss of generality, the stiffness values of 3 springs connected to the same node is assumed to be the same. Therefore, in this case analysis we have 4 spring stiffness values to be updated, i.e., the total number of unknowns to be updated is 4. Before conducting the model updating analysis, we first estimate the order of magnitude of such spring elements. It is found that once we increase the boundary stiffness values of the aforementioned 4 nodes to $1\times 10^{24}$ N/m, the natural frequencies of the dome structure approach those of the dome with all 18 nodes fixed. Therefore, we can consider $1\times 10^{24}$ N/m being the stiffness value corresponding to the fixed boundary condition. As such, we define the entire search bound as $[0, 1\times 10^{24}]$ N/m. This can ensure the detection of the response with respect to the boundary condition change. For this particular case, we treat the upper bound of stiffness (i.e., $1\times 10^{24}$ N/m) as the nominal value, and assume the actual spring stiffness reduction due to non-ideal boundary conditions at these 4 nodes are given as $\Delta \mathbf{K}=[0.7\times 10^{24}, 0.6\times 10^{24}, 0.1\times 10^{24}, 0.9\times 10^{24}]$ N/m. Therefore, the actual stiffness reduction coefficients to be identified are $\bar{\boldsymbol{\alpha}}=[0.7, 0.6, 0.1, 0.9]$. We define the prior PDF of model parameters, i.e., 4 spring stiffness reduction coefficients, as a multivariate uniform distribution with range [0, 1]. In what follows we use simulated data in lieu of experimental data to conduct the model updating practice. Since the 'ground truth', i.e., the true boundary condition, is known in this case study, we will be able to fully demonstrate the algorithmic improvement.

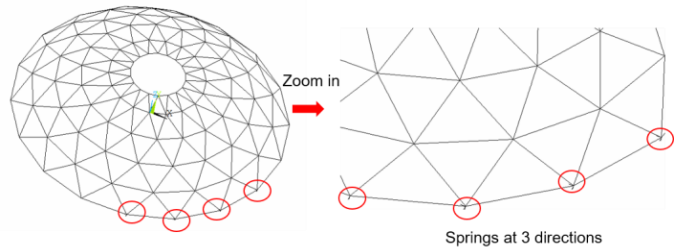

Figure 4. Boudary condtions (i.e., 4 spring stiffness values) of the dome structure to be updated.

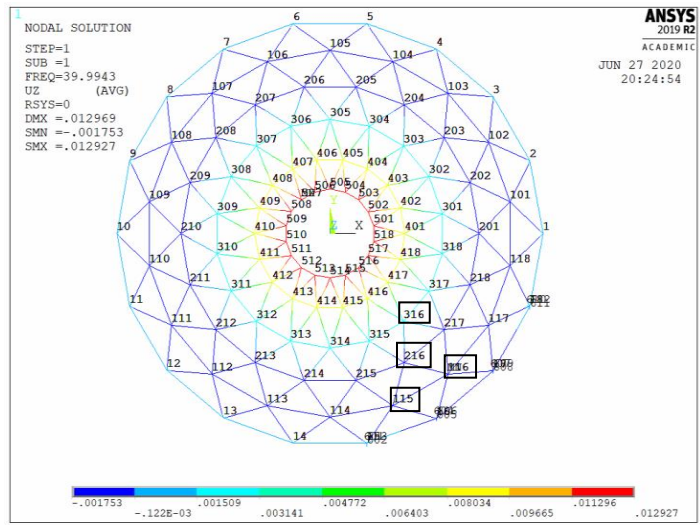

Figure 5. 1$^{st}$ $z$-bending mode shape of the dome structure attached with 4 sensors (locations: node 115, 215, 216 and 316).

Since we are subject to incomplete modal information measurement, we assume only the first two natural frequencies and the associated mode shapes are available for model updating. Moreover, we assume only 4 sensors (uniaxial accelerometers) installed along the $z$-direction at 4 nodes indicated in Figure 5 are employed. Therefore, only the amplitudes of the mode shapes at these DOFs are measurable. For demonstration, the 1$^{st}$ $z$-bending mode shape of the dome structure is also shown in Figure 5. Throughout model updating, the measurement information will be compared with model prediction continuously. Here we utilize the mode assurance criterion (MAC) to assess the difference between the mode shape prediction and the measurement. Given two mode shapes for comparison, MAC directly converts the vector difference into a scalar. As a result, the influence of mode shape and natural frequency can be equivalently incorporated into the formulation of posterior PDF. Recall that in Section 2.1 the differences of natural frequencies and mode shapes are generically expressed as $\kappa(\mathbf{\alpha})$ and $\upsilon(\mathbf{\alpha})$ where $\mathbf{\alpha}$ is the vector of model parameters to be updated. We now let

$$\kappa(\mathbf{\alpha}) = \sum_{i=1}^{q} \frac{|\hat{\omega}_i - \bar{\omega}_i|}{\bar{\omega}_i}, \qquad \upsilon(\mathbf{\alpha}) = \sum_{i=1}^{q} (1 - \gamma_i) \qquad (7a,b)$$

where $\gamma_i$ is MAC that is defined as $\gamma_i = \frac{\left|\hat{\mathbf{\psi}}_i^T \bar{\mathbf{\psi}}_i\right|^2}{\left(\hat{\mathbf{\psi}}_i^T \hat{\mathbf{\psi}}_i\right)\left(\bar{\mathbf{\psi}}_i^T \bar{\mathbf{\psi}}_i\right)}$ [40]. Hereafter the hat notation indicates variable obtained through model prediction, and the bar notation indicates variable obtained from measurement. As simulated data are used in lieu of actual measurement, $\bar{\omega}_i$ and $\bar{\mathbf{\psi}}_i$ are obtained numerically from the baseline model with $\bar{\mathbf{\alpha}} = [0.7, 0.6, 0.1, 0.9]$.

We further derive the likelihood PDFs of both $\kappa(\mathbf{\alpha})$ and $\upsilon(\mathbf{\alpha})$ with the normal distribution given as.

$$p(\kappa(\mathbf{\alpha})|\mathbf{\alpha}) = \prod_{i=1}^{q} \exp(\frac{-(|\hat{\omega}_i - \bar{\omega}_i|)^2}{2(\eta_i \bar{\omega}_i)^2}), \qquad p(\upsilon(\mathbf{\alpha})|\mathbf{\alpha}) = \prod_{i=1}^{q} \exp(\frac{-(1-\gamma_i)^2}{2(\eta_i)^2}) \qquad (8a,b)$$

A very important aspect of model updating problem is the measurement noise, as actual measurement is always subject to noise effect. It is worth emphasizing that the above likelihood PDFs take into account measurement noise through incorporating the variability level, i.e., $\eta_i$. The values of $\eta_i$ are generally decided based on the noise level of experimental measurements. Previous literature have reported that the values of $\eta_i$ can be estimated from the convergence of the statistics of the measured data, i.e., identified modal parameters. To allow unbiased estimation, usually a large number of repetitive measurements collected in the structure are needed [41]. These values indeed are case specific, depending on the structure investigated, data acquisition equipment resolution, sensor locations, ambient noise and mode order/frequency range of interest. In literature, the variability of measured natural frequencies generally falls into the range [0.5%, 3%] [26, 41-43]. It should be noted that modeling error could be more significant than measurement noise in practical scenarios [24]. To adequately take into account all the uncertainties for the posterior PDF characterization, relatively large values of $\eta_i$ are suggested. Therefore, in this research $\eta_i$ are chosen as 0.08 (i.e., 8%) for both the natural frequencies and MACs to take into account the noise/uncertainties. We assume the PDFs of natural frequencies and mode shapes are statistically independent. The final likelihood thus can be written as

$$p(\kappa(\mathbf{\alpha}), \upsilon(\mathbf{\alpha})|\mathbf{\alpha}) = \prod_{i=1}^{q} \exp(\frac{-(|\hat{\omega}_i - \bar{\omega}_i|)^2}{2(\eta_i \bar{\omega}_i)^2} + \frac{-(1-\gamma_i)^2}{2(\eta_i)^2}) \qquad (9)$$

Substituting Equation (9) into Equation (6) yields the close-form of the posterior PDF.

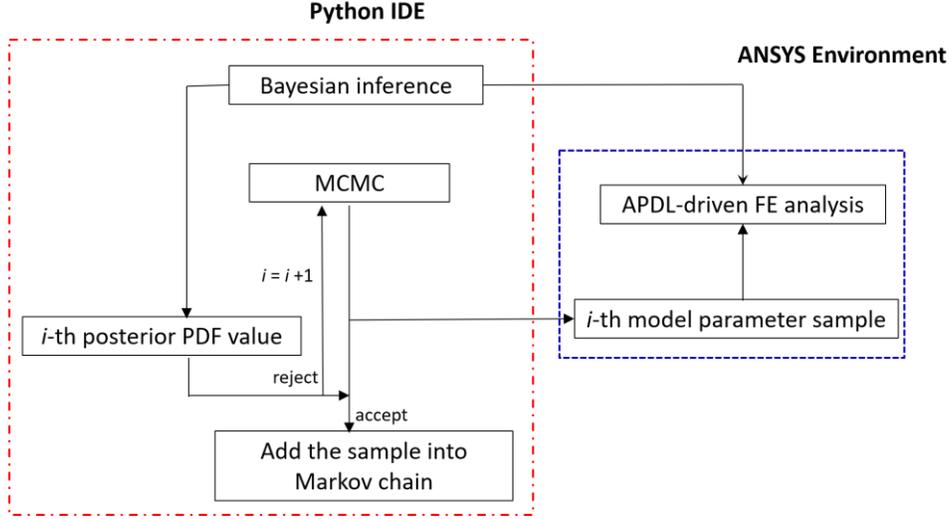

Figure 6. Analysis flowchart that enables the interaction between ANSYS and Python.

In this research, we establish the probabilistic updating framework using Python. We leverage the efficient solver of ANSYS [44] for finite element simulations, and incorporate it into the framework. APDL script is employed to direct the ANSYS analysis in the backstage. The interface between python IDE and ANSYS environment is built to facilitate the updating process. The framework developed is completely automated, and the entire Markov chain will be eventually produced through FE analysis iterations. For illustration, Figure 6 shows the architecture of conventional Bayesian model updating, indicating how the interaction between Python IDE and ANSYS environment takes place. This analysis structure can be further extended to suit the enhanced updating framework with multiple Markov chains. The APDL pseudo code for finite element analysis under certain model parameter sample is given in Appendix for interested readers to reproduce the analysis result.

*3.1.2. Model updating practice using incomplete modal information*

3.1.2.1. Direct parametric estimation with MCMC evolution

Following the procedures outlined in the Section 2.3, we first formulate a multivariate uniform distribution with bound [0,1] to facilitate the Latin hypercube sampling [39] of 1,000 (i.e., $w = 1,000$) model parameter samples. The posterior PDF values of these samples can be directly calculated using Monte Carlo simulation. We then screen 50 out of these 1,000 samples (i.e., $r = 50$) with higher posterior PDF values. With a pre-specified distance threshold, i.e., $\varepsilon_C = 0.23$, $K$-means clustering analysis is carried out to partition 50 pre-screened samples into 14 clusters. Each cluster is fully differentiated by its center information. The spatial coordinates of these centers are considered as the initial model parameters of respective 14 Markov chains to be executed.

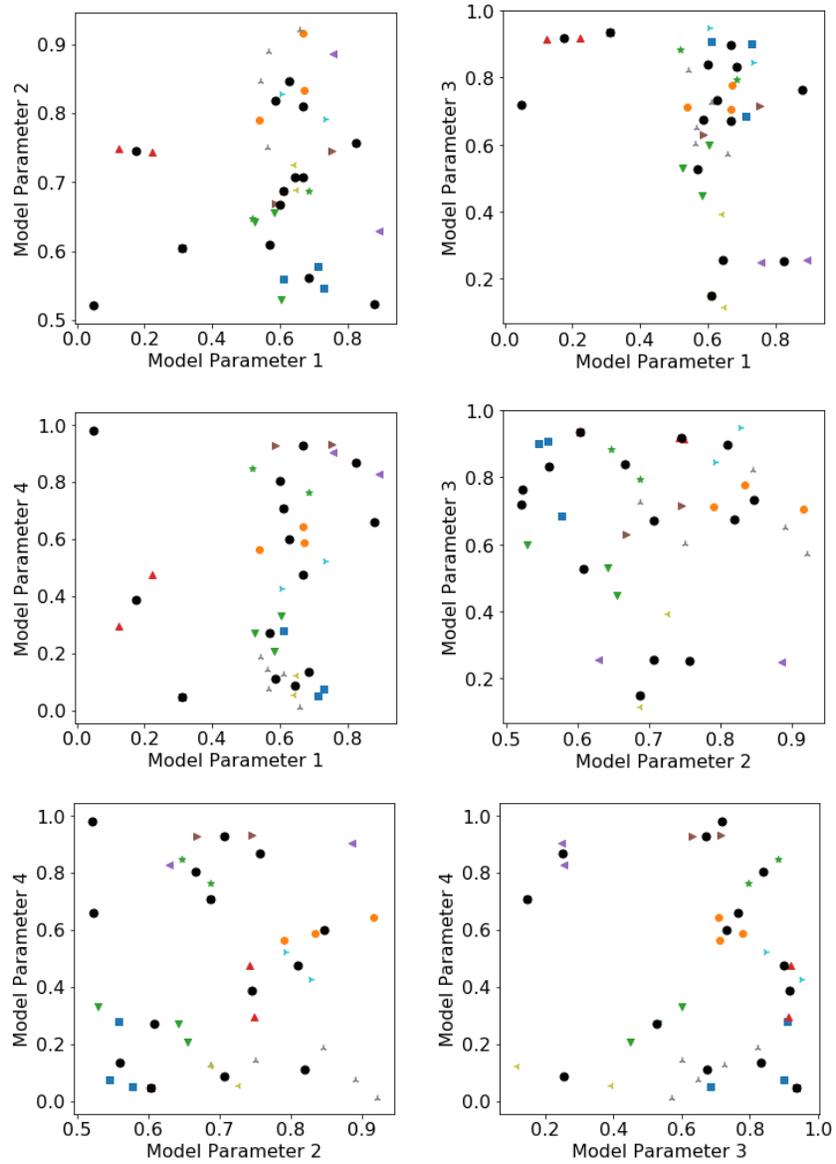

Figure 7. Initial model parameter samples using *K*-means clustering analysis based upon pre-screened ranked model parameter samples from Monte Carlo. (Circles denote the centers of clusters; other shapes denote the clusters.)

Figure 7 gives the 2-D cluster information yielded by *K*-means clustering analysis. The details of the operating variables can be found in Table 1. 14 Markov chains progressively evolve upon the initial parameters. Eventually, 6 Markov chains among them survive, and the rest of 8 merge to others. From the aspect of computational efficiency, we summarize the status of Markov chains as shown in Figure 8. In each iteration, the major computational costs are spent in FE analysis, which takes around 2-3 seconds on a desktop computer with Intel CPU E5-2640 @2.40GHz (2 processors). The overall computational cost therefore can be estimated in terms of the numbers of iterations for both merged and survived Markov chains (Figure 8).

Table 1. MCMC Parameters (Case 1)

| | |
|---|---|
| $w$: number of samples for Monte Carlo | 1,000 |
| $r$: number of best model parameter samples obtained through MC | 50 |
| $t$: pre-specified maximum number of simulation runs for each MCMC | 2,000 |
| $\varepsilon_C$ : distance threshold for determining initial parameter samples | 0.23 |
| $\varepsilon_M$ : distance threshold for merge check of Markov chains | 0.15 |
| $\gamma$ : default proposal distribution width | 0.01 |
| $u$: number of consecutive rejected steps for earlier termination | 250 |
| $\kappa$ : burn-in length ratio | 0.1 |

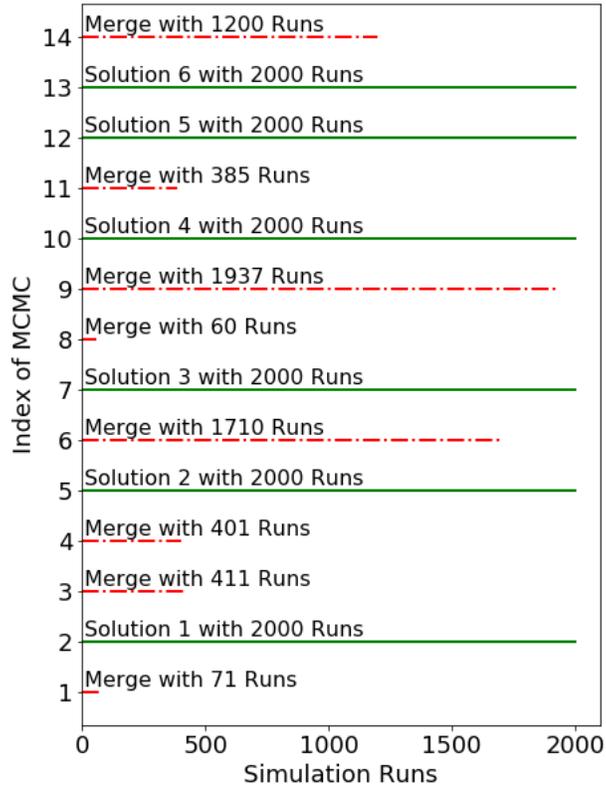

Figure 8. MCMC evolution progresses.

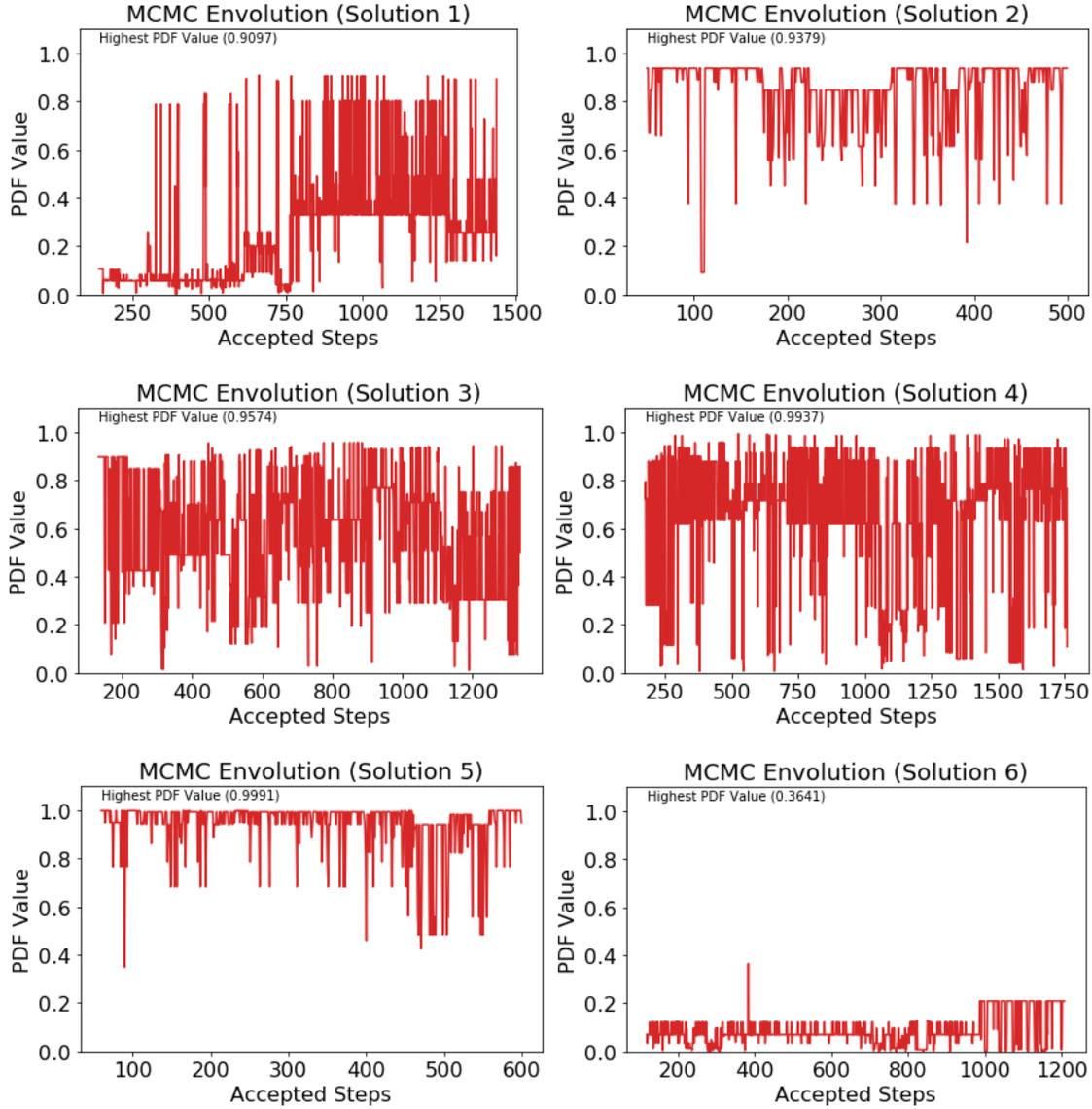

Figure 9. MCMC evolution and optimization.

Recall the pseudo code of MH MCMC shown in Section 2.2.2. An important procedure, i.e., sample acceptance in MCMC, collects the 'useful' samples to approximate the target distribution in a statistical manner. The complete sample acceptance history can be represented by the resulting Markov chain. The trends of posterior PDF values in survived Markov chains are presented in Figure 9. It is worth noting here one Markov time step in the horizontal axis denotes one accepted sample. For notation convenience, each survived Markov chain can be deemed as one solution. One may notice that the index of Markov time step does not start from 1 because of the removal of 'burn-in' period as mentioned before. The maximum iteration number of MCMC is specified as 2,000 (Table 1). The numbers of accepted time steps vary with respect to the Markov chain, which may be due to the different input-output relations around different local optima. Unlike general optimization methods, the objective value in this framework does not monotonically increase as process proceeds. This observation is directly due to the Metropolis criterion which takes place in the acceptance step (please refer to the MH MCMC pseudo code in Section 2.2.2). The random number generated for comparison will oscillate the objective values of accepted samples, which to certain extent can alleviate the trap of local optima. The maximum objective value is 1 as the posterior PDF is normalized. It is observed that the highest objective values of all solutions are relatively large. The values in Solutions 3, 4 and 5 even approach 1. The result illustrates the good performance of parameter search upon the MCMC evolution.

We now select the solutions with the highest objective values that are greater than 0.9 (i.e., Solutions 1-5), and compare their respective best model parameters as shown in Figure 10. Apparently, model parameter values have noticeable discrepancy, which indicates the existence of multiple local optima in the objective surface. The new framework developed is indeed capable of capturing the underlying information of these local optima.

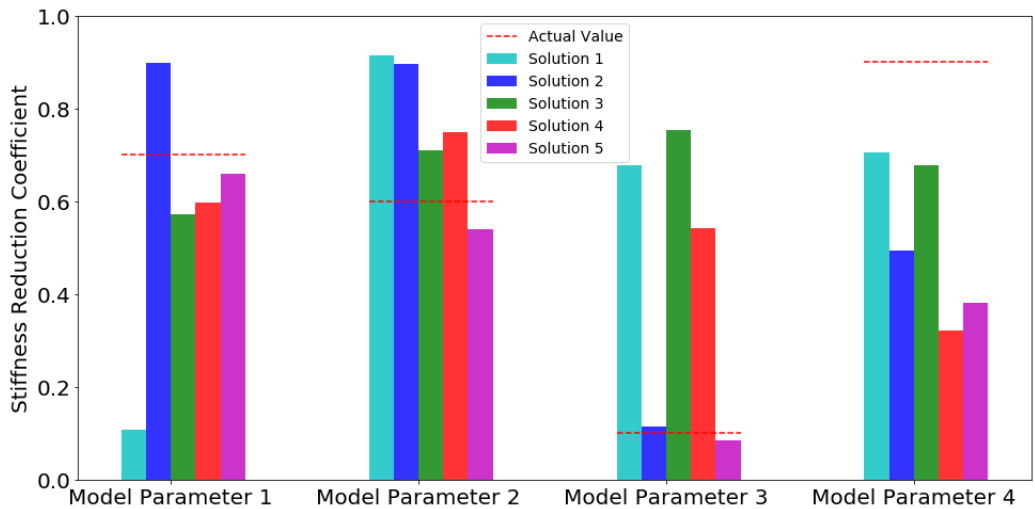

Figure 10. Directly identified model parameters from the best 5 solutions through MCMC evolution.

3.1.2.2. Target function approximation and further parameter identification

While the best model parameters identified directly from the abovementioned solutions indeed result in extremely high objective value, such as Solutions 4 (0.9937) and 5 (0.9991), they cannot be considered as final updating result. The reason is that a small number of data points, i.e., accepted samples and their objective values, cannot well characterize the true posterior PDF (Equation (6)) especially when the parametric space is high-dimensional. To further enhance the accuracy of updating result, previous research [10] established a meta-model based on the scarce posterior PDF data points from MCMC, and then employed it to enrich the posterior PDF. The parameter estimation can then be conducted upon this enriched posterior PDF. It is worth pointing out that the underlying idea of MCMC actually allows an alternative for parameter optimization. Specifically, MCMC aims at constructing Markov chain that has desired distribution with respect to the target function/distribution [39]. The information of Markov chain in fact is fully represented by the accepted model parameter samples without the respective objective values. Therefore, the approximated distribution essentially is a histogram, representing the frequency of occurrence of the model parameter combination.

A direct way to convert Markov chain into a histogram is briefly introduced as follows. For example, consider the $i$-th solution/survived Markov chain with recorded information $\mathbf{\Theta}_{h \times g}^{(i)}$, where $h$ is the length of the chain, and $g$ is the number of model parameters to be updated, i.e., $g = 4$ in this current case. One can slice $\mathbf{\Theta}_{h \times g}^{(i)}$ into $g$ column vectors, which then can be presented with histograms. In statistics, each histogram in the $i$-th solution is used to represent the marginal PDF of certain model parameter, e.g., a marginal PDF denoted as $p(\mathbf{\alpha}_j)$, where $\mathbf{\alpha}_j$ is the $j$-th model parameter. In this case, we partition the entire range of $\mathbf{\alpha}_j$, i.e., [0, 1] into 20 uniform small bins, upon which the marginal PDFs of all model parameters are generated. While the statistical properties of such PDFs indeed indicate the probabilistic identification result, they cannot be directly utilized for identifying the best model parameters. The reason is that they are characterized based upon the assumption that all model parameters are independent. However, different model parameters essentially are coupled when characterizing the posterior PDF. The lack of coupling causes the inconsistency of frequency of occurrence at the same solution. To clarify this, we take Solution 1 as an example (Figure 11). Obviously, the highest frequencies of occurrence of different model parameters are not identical.

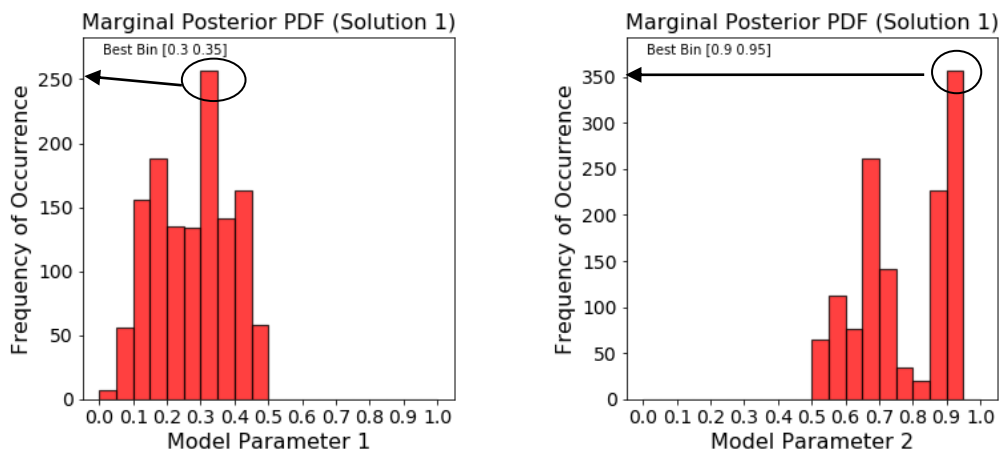

Figure 11. Inconsistency of highest frequency of occurrence in marginal posterior PDF.

Theoretically, the joint PDF is a quantity that can truly reflect the probability of model parameter combination to be the actual one. We thus use the joint PDF instead of the marginal PDF to conduct the subsequent probabilistic parameter identification. The mathematical relation between joint and marginal PDFs can be characterized as

$$p(\pmb{\alpha}_j) = \int_{\vee \pmb{\alpha}_1} .. \int_{\vee \pmb{\alpha}_k (k \neq j)} .. \int_{\vee \pmb{\alpha}_n} p(\pmb{\alpha}_1,...,\pmb{\alpha}_j...\pmb{\alpha}_n) d\pmb{\alpha}_1 .. d\pmb{\alpha}_k .. d\pmb{\alpha}_n \tag{11}$$

where $p(\pmb{\alpha}_1,...,\pmb{\alpha}_j...\pmb{\alpha}_n)$ is the joint PDF. According to the above equation, the marginal PDF of the *j*-th model parameter is calculated as a multiple integral of joint PDF over the entire high-dimensional parametric space without the *j*-th dimension. For the sake of computation, we oftentimes approximate such integral using a discrete form, expressed as

$$p(\pmb{\alpha}_j) = \sum_{\vee \pmb{\alpha}_1} .. \sum_{\vee \pmb{\alpha}_k (k \neq j)} .. \sum_{\vee \pmb{\alpha}_n} p(\pmb{\alpha}_1,...,\pmb{\alpha}_j...\pmb{\alpha}_n) \Delta \pmb{\alpha}_1 .. \Delta \pmb{\alpha}_k .. \Delta \pmb{\alpha}_n \tag{12}$$

It is noteworthy that the characterization of joint PDFs using histogram plots requires quite high computational cost, because the joint PDFs are multidimensional. The data to be used to create the histogram plots include the specified bins and relevant count numbers of samples fall within those bins. Assuming that we divide $Q$ bins within the range [0, 1] for all 4 model parameters, the total number of bins will be $Q^4$. When choosing a large value of $Q$ to reflect the continuous property of model parameters, the sample count procedure will be quite costly in order to loop over all $Q^4$ bins. To facilitate this counting procedure, in this research we set a small number of $Q$ as 20. According to the chain evolution shown in Figure 9, Solution 6 certainly is not a potential local optimum. As such, we only produce the joint PDFs of the other 5 Markov chains, as shown in Figures 12-16. The multidimensional histogram plots are projected into different low-dimensional input spaces for the sake of visualization. Particularly, the final projected joint PDF of one model parameter is the overlay of $20^{(4-1)}$ projected PDFs built upon the other different model parameter combinations. For illustration purpose, in each histogram plot we only add 5 non-zeros respective projected PDFs instead of total $20^{(4-1)}$ projected PDFs. As a comparison, we also add the projected PDF with the highest frequency of occurrence value identified, from which we can easily find the corresponding best model parameter. This new set of results not only provides a rational mean for parameter identification, but also probabilistically interprets the result when the measurement and modeling uncertainties are taken into account. For example, the confidence level of updating result is given.

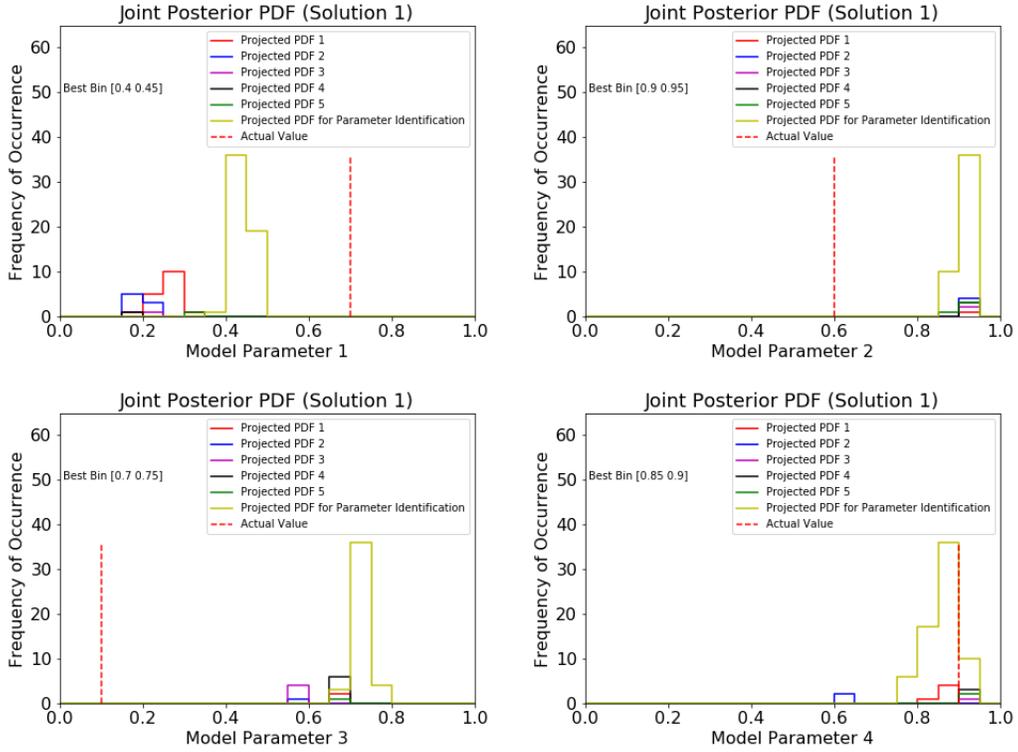

Figure 12. Joint PDF constructed from Solution 1.

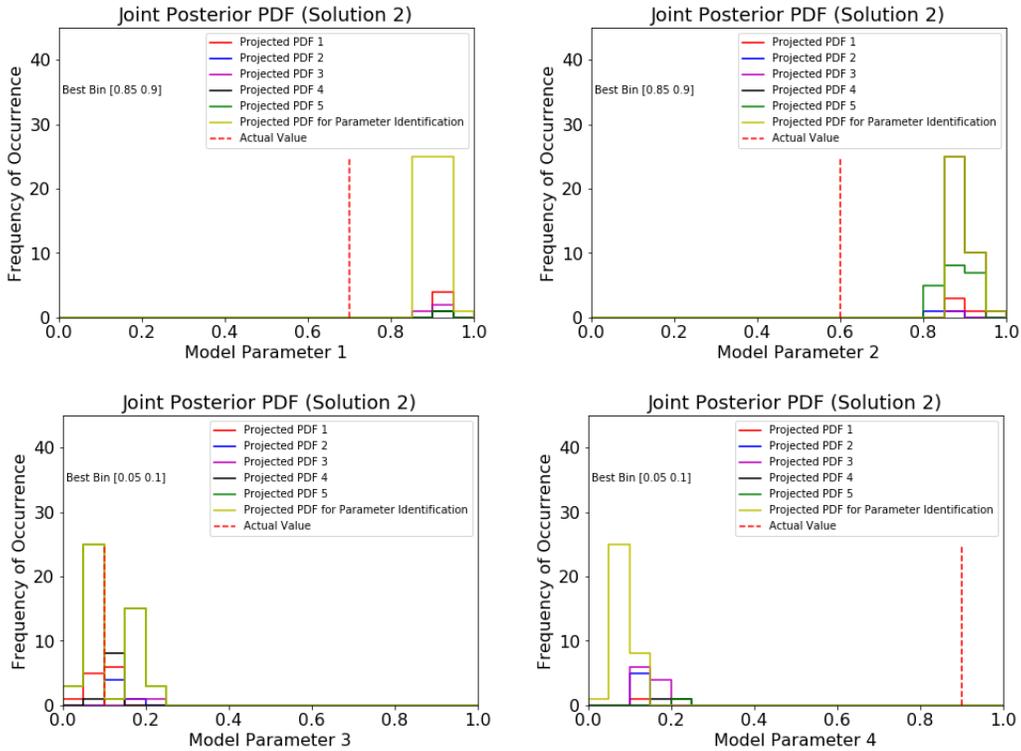

Figure 13. Joint PDF constructed from Solution 2.

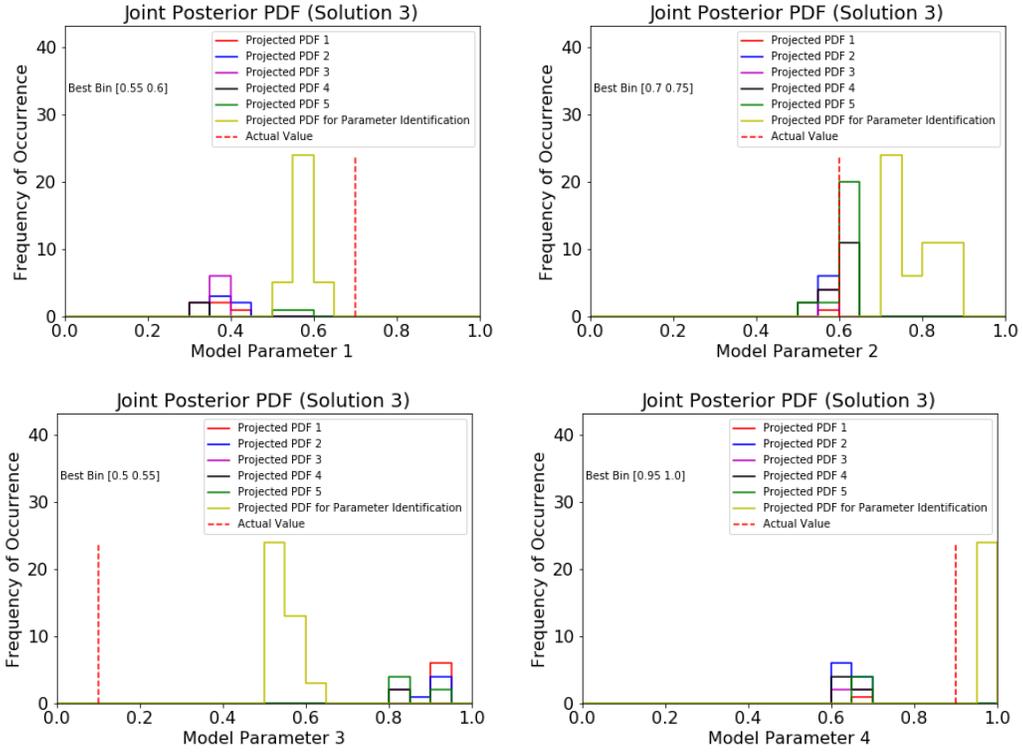

Figure 14. Joint PDF constructed from Solution 3.

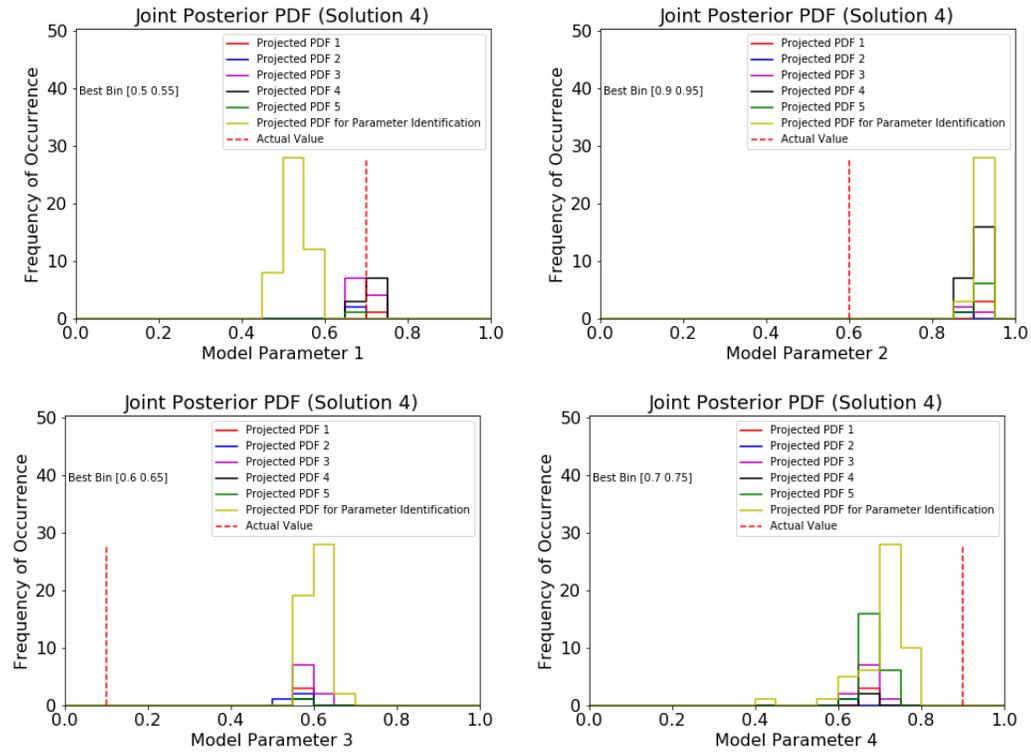

Figure 15. Joint PDF constructed from Solution 4.

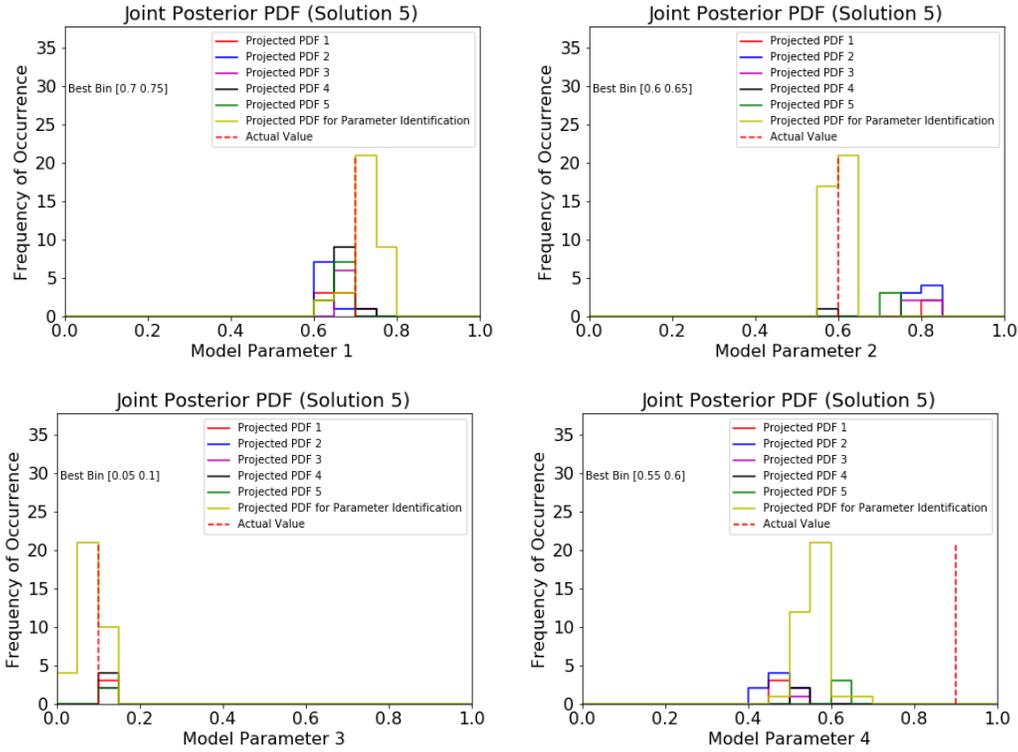

Figure 16. Joint PDF constructed from Solution 5.

The actual model parameters (i.e., the ground truth) are represented as vertical dash line in the figures for comparison. The bins covering the best model parameters among those solutions have obvious variation, which again reflects the fact that the multiple local optima exist in this model updating problem. Compared with other solutions, the bins identified in Solution 5 are closer to the values of actual model parameters. Specifically, the identified bins of model parameters 1, 2, 3 exactly match the respective actual model parameters. Only the identified bin of model parameter 4 has certain discrepancy with respect to the actual value. From physical perspective, model parameter 4 may indeed be insensitive to the selected modal responses.

Overall, the results clearly illustrate the capability of this framework to probabilistically identify multiple local optima. This in fact is advantageous in actual model updating practice, in which the actual stiffness values at boundaries are unknown. The multiple solution options acquired can allow us approach the ground truth from various angles. One way of finalizing the solution option is to employ empirical knowledge and experience. For example, if we're only interested in response within certain frequency range, the best solution would be the one that yields the minimum difference between measurement and model prediction within that frequency range. In certain situations, we may gather additional information (e.g., additional modal information or additional sensor) to assist decision making.

*3.2. Implementation scenario 2: Stiffness reduction identification in plate structure*

We then analyze the second implementation scenario, i.e., identification of stiffness reduction in a plate structure with large number of DOFs. As shown in Figure 17a, a plate structure with dimensions $0.4 \times 0.4 \times 0.005$ (m) is investigated. It is clamped at two edges along the *x*-axis. The material constants are: Young's modulus $2.06 \times 10^{11}$ Pa, mass density $7.85 \times 10^3$ kg/m$^3$, and Poisson's ratio $0.3$. The plate is meshed with 8-node solid element within ANSYS. The total number of DOFs is 10,086. We divide the plate into 8 uniform segments along the *x*-axis, and our goal is to identify the change/reduction of stiffness in these segments, possibly caused by damage or material property non-uniformity. Therefore, this second implementation scenario applies to structural damage detection or model calibration with material property updating. We assign one stiffness reduction coefficient to each segment, so altogether we have 8 parameters to update, i.e., $n = 8$ in Equation (2). In this simulated case, we assume the actual stiffness reduction coefficients, i.e., the 'ground truth', are known as $\bar{\boldsymbol{\alpha}}=[0.2, 0.5, 0.6, 0.1, 0.6, 0.3, 0.2, 0.7]$. We assume only the information of the first two *z*-direction bending mode shapes is available. Moreover, we assume only 4 sensors are employed, so for the first two modes, only the amplitudes at 4 DOFs are measured. This leads to a model updating problem with severely limited and incomplete measurement. This scenario is considerably different from the first scenario as the number of DOFs in the baseline model is very large. Coupled with the incomplete measurement, the objective surface in Bayesian inference based model updating is very complicated. Figure 17b illustrates the first *z*-direction bending mode shape of the structure, and shows 4 sensors that are uniformly distributed along the *x*-axis.

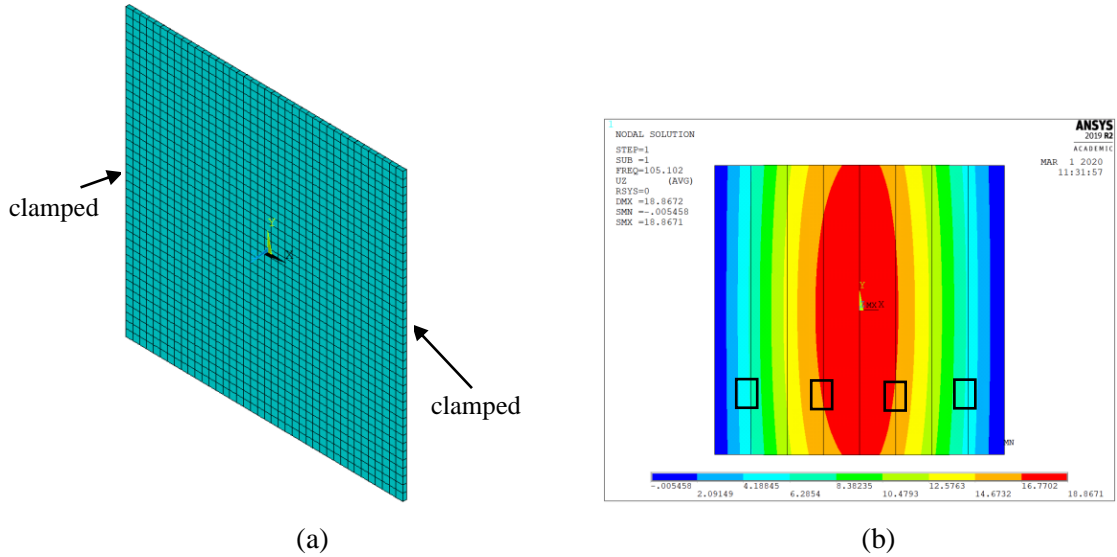

(a)          (b)

Figure 17. Model updating setup (a) FE model; (b) 1st $z$-direction bending mode shape contour plot of structure with 8-segment stiffness variations and 4 selected measurement locations (denoted by square).

Similar to implementation scenario 1, in this second case a uniform distribution within the entire parametric space, i.e., [0, 1], is specified to characterize the prior PDF. Since in this second case we intend to utilize solely two incomplete bending mode shapes for model updating, we therefore only involve mode shape difference $\upsilon(\boldsymbol{\alpha})$ in the posterior PDF derivation following Equation (6). It is worth noting that there indeed exist various ways of describing the difference between model prediction and measurement. In this second case, we adopt a new expression of mode shape difference, i.e., point-to point mode shape amplitude differences. The reason is that this can fully take advantage of the limited measurement information.

$$\upsilon(\boldsymbol{\alpha}) = \begin{bmatrix} \psi_{1,1} & \psi_{1,2} & \cdots & \psi_{1,q} \\ \psi_{2,1} & \psi_{2,2} & \cdots & \psi_{2,q} \\ \cdots & \cdots & \cdots & \cdots \\ \psi_{s,1} & \psi_{s,2} & \cdots & \psi_{s,q} \end{bmatrix} - \begin{bmatrix} \bar{\psi}_{1,1} & \bar{\psi}_{1,2} & \cdots & \bar{\psi}_{1,q} \\ \bar{\psi}_{2,1} & \bar{\psi}_{2,2} & \cdots & \bar{\psi}_{2,q} \\ \cdots & \cdots & \cdots & \cdots \\ \bar{\psi}_{s,1} & \bar{\psi}_{s,2} & \cdots & \bar{\psi}_{s,q} \end{bmatrix} = \begin{bmatrix} \Delta\psi_{1,1} & \Delta\psi_{1,2} & \cdots & \Delta\psi_{1,q} \\ \Delta\psi_{2,1} & \Delta\psi_{2,2} & \cdots & \Delta\psi_{2,q} \\ \cdots & \cdots & \cdots & \cdots \\ \Delta\psi_{s,1} & \Delta\psi_{s,2} & \cdots & \Delta\psi_{s,q} \end{bmatrix} \quad (10)$$

where $q$ is the number of mode shapes and $s$ is the number of measurement locations. To take into account the effect of measurement noise and modeling error, in this case the individual likelihood PDF for each mode shape amplitude is formulated as a normal distribution, given as

$$p(\Delta\psi_{j,k} | \boldsymbol{\alpha}) = \exp\left(\frac{-(\Delta\psi_{j,k})^2}{2(\eta_{j,k}\bar{\psi}_{j,k})^2}\right) \quad (11)$$

where $\eta_{j,k}$ is the variance, indicating the uncertainty degree of the actual mode shape measurement $\bar{\psi}_{j,k}$. Here we set $\eta_{j,k}$ as 0.1 in this case. As discussed in Section 3.1.1, this value can adequately account for all the uncertainties and measurement noise. This likelihood PDF can ensure the higher probability of $\boldsymbol{\alpha}$ to be the actual model parameters when a smaller $\Delta\psi_{j,k}$ is observed. Since all elements in $\upsilon(\boldsymbol{\alpha})$ have similar effect, we can write a final likelihood PDF in a multiplication form as

$$p(\upsilon(\boldsymbol{\alpha}) | \boldsymbol{\alpha}) = \prod_{j=1}^{s}\prod_{k=1}^{q} p(\Delta\psi_{j,k} | \boldsymbol{\alpha}) \quad (12)$$

Following the procedure outlined in Section 2, we can obtain the posterior PDF.

Once again, we specify the MCMC parameters and carry out the numerical analysis. Here it is worth nothing that we select a larger threshold value, i.e., $\varepsilon_C = 0.35$ than that of case 1 for clustering analysis because the dimension of parametric space in this case becomes higher. As a result, the spatial distance of different clusters generally will increase. In addition, a larger maximum number of simulation runs of MCMC, i.e., $t = 3,000$ is adopted to enable the chain evolution convergence because of higher dimensional FE model and higher dimensional parametric space in this updating problem. Other parameters are kept the same as shown in Table 1.

By checking the Markov chain evolution history without the 'burn-in' period, all survived chains are supposed to point to the different local optima because the associated highest PDF values are greater or around 0.9. In this case, a reduced number of bins, i.e., 10 is defined to divide each 1-dimensional parametric space, since the counting over total $10^8$ bins is already costly. The joint posterior PDFs calculated indicate the probabilistic updating results, in which the best parameter bins can be identified and summarized in Table 2. The solution highlighted in bold font is very close to the actual values. Specifically, the identified bins of model parameters 1, 2, 3, 4, 6, 8 exactly encompass the respective actual values. Only the identified bin of model parameter 5 has certain deviation with respect to the actual value. Moreover, while Solution 1 outperforms Solution 6 in terms of the highest objective value comparison, it appears to point to another local optimum where measurement noise may play a role. Intuitively, the model parameters 7 and 8 are more difficult to be identified accurately, which may be due to their low sensitivity to the selected modal responses. Overall, the result indicates the consistent good updating performance of this new method. Once again, multiple solution options here allow us to carry out decision making possibly using additional information, e.g., NDE in structural fault identification, to pinpoint the root cause.

**Table 2.** Identified stiffness coefficient bin based upon the joint posterior PDF of MCMC (Case 2)

|         | Para. 1    | Para. 2    | Para. 3    | Para. 4    | Para. 5    | Para. 6    | Para. 7    | Para. 8    | Highest PDF |
|---------|------------|------------|------------|------------|------------|------------|------------|------------|-------------|
| Solu. 1 | [0.3, 0.4] | [0.6, 0.7] | [0.6, 0.7] | [0.2, 0.3] | [0.5, 0.6] | [0.2, 0.3] | [0.6, 0.7] | [0.1, 0.2] | 0.9809      |
| Solu. 2 | [0.3, 0.4] | [0.2, 0.3] | [0.3, 0.4] | [0.7, 0.8] | [0, 0.1]   | [0, 0.1]   | [0.4, 0.5] | [0, 0.1]   | 0.9111      |
| Solu. 3 | [0, 0.1]   | [0.3, 0.4] | [0.2, 0.3] | [0.4, 0.5] | [0.8, 0.9] | [0, 0.1]   | [0.5, 0.6] | [0.6, 0.7] | 0.9554      |
| Solu. 4 | [0.1, 0.2] | [0.1, 0.2] | [0.5, 0.6] | [0.3, 0.4] | [0, 0.1]   | [0.5, 0.6] | [0.1, 0.2] | [0.6, 0.7] | 0.8733      |
| Solu. 5 | [0.4, 0.5] | [0.2, 0.3] | [0.5, 0.6] | [0, 0.1]   | [0.1, 0.2] | [0.2, 0.3] | [0.4, 0.5] | [0, 0.1]   | 0.9024      |
| **Solu. 6** | **[0.1, 0.2]** | **[0.5, 0.6]** | **[0.6, 0.7]** | **[0.1, 0.2]** | **[0.2, 0.3]** | **[0.3, 0.4]** | **[0, 0.1]** | **[0.7, 0.8]** | **0.9688** |
| Actual  | 0.2        | 0.5        | 0.6        | 0.1        | 0.6        | 0.3        | 0.2        | 0.7        |             |

The identified bins in the highlighted solution is the closest to the actual values.

## 4. Conclusions

This paper presents a new model updating framework using incomplete measurement in the presence of uncertainties. The framework is established upon the Bayesian inference through conducting parameter updating in terms of the posterior PDF. With limited measurement, multiple local optima likely exist in the parametric space. To tackle the issue, we synthesize an enhanced Bayesian approach by incorporating multiple parallel, interactive and adaptive Markov chains. The joint posterior PDFs constructed by the final survived Markov chains can be used to interpret the probabilistic updating results. We carry out the systematic case investigations through formulating different model updating problems, i.e., boundary and material property updating of a dome and a plate structures, respectively. The results indicate that multiple optima can indeed be identified in terms of the joint posterior PDF computed via this numerical framework, and the 'ground truth' is included in the solution set with high PDFs. The statistical features of posterior PDF also indicate the confidence level of parameter estimation in the presence of uncertainties. This approach can be applied to a variety of model updating and fault identification applications.

**Acknowledgment**

This research is supported in part by a Space Technology Research Institutes grant (number 80NSSC19K1076) from NASA's Space Technology Research Grants Program, and in part by NSF under grant CMMI-1825324.

**Appendix: Pesudo Code of ANSYS APDL**

The APDL pseudo code for finite element analysis under certain model parameter sample is given as follows.

| |
|---|
| *Resume the baseline model (constructed beforehand)* |
| RESUME,'dome_modal','db','D:\...........', 0,0 |
| *Read model parameter sample from intermediate file generated by external optimization code* |
| *DIM,unSamp,,1,n_inputs |
| *VREAD,UnSamp(1,1),input_data,txt,,JIK,8,1 |
| (8F8.4) |
| *Modify model via changing stiffness values of springs at the boundaries in terms of new model parameter sample* |
| /PREP7 |
| *DO,JJ,1, n_inputs |
| R,JJ+1,(1-Unsamp(1,JJ))*10e24,0,0, , ,0, |
| *ENDDO |
| *Solve the analysis* |
| NSEL, ALL |
| FINISH |

```
/SOL
SOLVE
FINISH
```

*Extract modal responses from defined measurement locations and write into intermediate file for optimization code*

```
*DIM,NODEIDT,ARRAY,1,n_locations
*DIM MODE,ARRAY,n_modes,,n_locations
*DIM MO,ARRAY,n_modes,1
*DIM,FREQ,ARRAY,1, n_modes
…
/POST1
*DO,I1,1, n_modes
*GET,FREQ(1,I1),MODE,SELORD(I1),FREQ
*ENDDO
*MWRITE,FREQ,output_Freq,TXT,,JIK,n_modes,1
(2F9.5)
*DO,I,1,n_modes
SET,,, ,,, ,MO(I)
*DO,I1,1,n_locations
*GET, MODE(I,I1),NODE,NODEIDT(I1,1),U,Z
*ENDDO
*ENDDO
*MWRITE,MODE,output_Mode,TXT,,JIK,n_locations,n_modes
(4F13.5)
```

Note: all the deterministic variables, such as the geometry, mesh and fixed boundaries of some nodes etc, have been set up in the baseline model. Therefore, in the model updating APDL script, there is no need to re-define those variables.